\documentclass[aps,preprint,amssymb,superscriptaddress]{revtex4-1}
\usepackage{floatrow}
\newfloatcommand{capbtabbox}{table}[\FBwidth]
\usepackage{graphicx}
\usepackage{dcolumn}
\usepackage{bm}
\usepackage{natbib}
\usepackage{CJK}
\usepackage[utf8]{inputenc}
\usepackage[T1]{fontenc}
\usepackage{mathptmx}
\usepackage{gensymb}
\usepackage{graphicx}
\usepackage{epstopdf}
\usepackage{dcolumn}
\usepackage{bm}
\usepackage{amsmath}
\usepackage{mathtools}
\usepackage{relsize}
\usepackage{multirow}
\usepackage[colorlinks]{hyperref}
\begin{document}

\title{Structure-Adaptive Topology Optimization Framework for Photonic Band Gaps with TE-Polarized Sources}

\author{Sukhad  Dnyanesh Joshi}
\affiliation{Department of Electrical Engineering and Computer Science, Syracuse University, Syracuse, NY 13210}
\author{Aditya Bahulikar}
\affiliation{Department of Electrical Engineering and Computer Science, Syracuse University, Syracuse, NY 13210}
\author{Feng Wang}
\affiliation{Department of Electrical Engineering and Computer Science, Syracuse University, Syracuse, NY 13210}
\author{M. Cenk Gursoy}
\affiliation{Department of Electrical Engineering and Computer Science, Syracuse University, Syracuse, NY 13210}
\author{Rodrick Kuate Defo}
\email{rkuatede@syr.edu}
\affiliation{Department of Electrical Engineering and Computer Science, Syracuse University, Syracuse, NY 13210}

\date{\today}

\begin{abstract}
Leveraging our structure-adaptive topology optimization framework based on the integration of the photonic density of states over a frequency window for the TM polarization of light [see A. Bahulikar~\textit{et al.}, arXiv:2411.09165 (2025)], we show that the $\Gamma$-point and full Brillouin zone integration schemes can also recover two-dimensional photonic crystals for TE polarization. For the $\Gamma$-point formalism, we employ the scalar magnetic field formulation of the electromagnetic wave equation with independent sources polarized in the $x$ and $y$ directions. For the full Brillouin zone formalism, we employ the vector electric field formulation of the electromagnetic wave equation, again with independent sources polarized in the $x$ and $y$ directions. This work can simultaneously treat frequency-dependent optical response, allow for targeted optimization for a given frequency and reciprocal lattice vector pair, and inherently encourage binarized designs.
\end{abstract}

\maketitle

\section{Introduction}
Integrated photonics is gaining traction due to the large bandwidth and low loss that photonic systems can accommodate~\cite{ou_hypermultiplexed_2025,agrawal_fiber_2012,winzer_fiber_2018,harris_high_2025,bernson_scalable_2025,buzzi_spectral_2025,greenspon_designs_2025,tao_single_2025,christen_an_2025}. The inverse design of nanophotonic systems has seen tremendous advances as well~\cite{hammond_unifying_2025,strekha_suppressing_2024,kim_automated_2023,chao_maximum_2022,chao_physical_2022,bahulikar_structure_2025,dory_inverse_2019,minkov_inverse_2020,deng_inverse_2024,kang_large_2024,takiguchi_inverse_2025,wang_inverse_2024,jensen_topology_2011,molesky_inverse_2018,li_topology_2019,sheverdin_photonic_2020} with investigations encompassing enhancement~\cite{chao_maximum_2022} and suppression~\cite{strekha_suppressing_2024} of the electromagnetic local density of states, engineering of photonic crystal cavities and waveguides~\cite{dory_inverse_2019,minkov_inverse_2020,takiguchi_inverse_2025}, design of photonic crystals~\cite{deng_inverse_2024,bahulikar_structure_2025,wang_inverse_2024}, and integration of neural networks in the design process~\cite{kang_large_2024}. As Takiguchi~\textit{et al}~\cite{takiguchi_inverse_2025} have noted, inverse design methods for nanophotonics can be classified as either heuristic~\cite{vasco_global_2021,goh_genetic_2007,wang_inverse_2024,takiguchi_inverse_2025}, gradient-based~\cite{minkov_inverse_2020,bahulikar_structure_2025,strekha_suppressing_2024,chao_maximum_2022}, or involving machine learning~\cite{asano_iterative_2019,li_deep_2023}. Gradient-based approaches provide reliable convergence and modest computational requirements, though at the cost of potentially missing globally optimal designs~\cite{takiguchi_inverse_2025}. Our approach alleviates the concern of missing globally optimal designs as it targets a specific photonic band gap, which is easy to verify, rather than simply seeking to maximize the band gap. 

Our approach proceeds by optimizing for specific target midgap frequencies and band gaps through minimization of the analytic continuation of the photonic density of states (DOS) over the appropriate frequency window. We leverage the formalism of Liang and Johnson~\cite{liang_formulation_2013} to cast the problem of minimization of the analytic continuation of the photonic DOS over the frequency window as minimization of a sum of evaluations at complex frequencies of the analytically continued photonic DOS~\cite{bahulikar_structure_2025}. Our approach uses a uniform source instead of point-dipole sources to reduce the computational cost of evaluating the photonic DOS by a factor of the number of grid points in the design region. Furthermore, our approach lends itself readily to the treatment of frequency-dependent optical response~\cite{bahulikar_structure_2025}, unlike approaches based on the solutions to eigenvalue equations. In this work, we treat the TE polarization of light and our work has potential applications in investigations of low-dimensional materials~\cite{mattheakis_epsilon_2016,kuate_methods_2021,kuate_strain_2016,shiang_ab_2015}. Work is forthcoming that treats complete photonic band gaps and three dimensions.

\section{Background and Approach}
\subsection{Deriving the photonic-crystal objective}
For the TE polarization of light, we consider both the magnetic- and electric-field wave equations with a source to obtain the photonic local density of states (LDOS). These wave equations have the form~\cite{men_robust_2014,liang_formulation_2013,bahulikar_structure_2025}
\begin{equation}
\label{eq:Maxwell}
\mathcal{M}^{\mathbf{H}}(\epsilon,\mu,\omega)\mathbf{H}_{j,\mathbf{r}'}(\omega,\mathbf{r}) = \mathbf{\nabla}\times\left(\frac{\mathbf{J}_{j,\mathbf{r}'}(\mathbf{r})}{\epsilon(\mathbf{r})}\right)
\end{equation}
and 
\begin{equation}
\label{eq:MaxwellE}
\mathcal{M}^{\mathbf{E}}(\epsilon,\mu,\omega)\mathbf{E}_{j,\mathbf{r}'}(\omega,\mathbf{r}) = i\omega\mathbf{J}_{j,\mathbf{r}'}(\mathbf{r}),
\end{equation}
where
\begin{equation}
\mathcal{M}^{\mathbf{H}}(\epsilon,\mu,\omega) = \mathbf{\nabla}\times\frac{1}{\epsilon(\mathbf{r})}\mathbf{\nabla}\times~ -~ \mu(\mathbf{r})\omega^2
\end{equation}
and
\begin{equation}
\label{eq:MaxwellopE}
\mathcal{M}^{\mathbf{E}}(\epsilon,\mu,\omega) = \mathbf{\nabla}\times\frac{1}{\mu(\mathbf{r})}\mathbf{\nabla}\times~ -~ \epsilon(\mathbf{r})\omega^2
\end{equation}
are the Maxwell operators for the magnetic and electric fields with $\mu(\mathbf{r})$ and $\epsilon(\mathbf{r})$ being the permeability and permittivity in the design region, respectively, and $\omega$ being a frequency. 
We note that when the magnetic-field wave equation is employed, the electric field can be obtained through
\begin{equation}
\mathbf{E}_{j,\mathbf{r}'}(\omega,\mathbf{r}) = -\frac{\mathbf{\nabla}\times\mathbf{H}_{j,\mathbf{r}'}(\omega,\mathbf{r})}{i\omega\epsilon(\mathbf{r})}.
\label{eq:HtoE}
\end{equation}
The quantity
\begin{equation}
\mathbf{J}_{j,\mathbf{r}'}(\mathbf{r}) = \delta(\mathbf{r}-\mathbf{r}')\hat{e}_j
\end{equation}
is the phasor representation of the time-dependent current source $\mathbf{J}_{j,\mathbf{r}'}(\mathbf{r},t) = \hat{e}_je^{-i\omega t}\delta\left(\mathbf{r}-\mathbf{r}'\right)$. We take this source to be a harmonic point-dipole source polarized in the direction of the unit vector $\hat{e}_j$ and with frequency $\omega$. The quantity $\delta\left(\mathbf{r}-\mathbf{r}'\right)$ is a Dirac delta function. Using the current source and corresponding electric field, we can then construct the per-polarization photonic local density of states (LDOS$_j\left(\omega,\mathbf{r}'\right)$), which captures the density of electric-field modes at a given point $\mathbf{r}'$ in space~\cite{oskooi_electromagnetic_2013}. Explicitly
~\cite{liang_formulation_2013,daguanno_electromagnetic_2004,novotny_principles_2012,inoue_photonic_2004,martin_electromagnetic_1998,taflove_advances_2013,bahulikar_structure_2025},
\begin{equation}
\label{eq:partialLDOS}
\text{LDOS}_j\left(\omega,\mathbf{r}'\right) = -\frac{6}{\pi}\text{Re}\left[\int\mathbf{J}_{j,\mathbf{r}'}^*(\mathbf{r})\cdot\mathbf{E}_{j,\mathbf{r}'}(\omega,\mathbf{r})\text{d}\mathbf{r}\right].
\end{equation}

Summing over polarizations and integrating over the entire crystal~\cite{taflove_advances_2013} yields the photonic density of states (DOS), which will be an important quantity in our optimization approach. As in prior work~\cite{liang_formulation_2013,shim_fundamental_2019,bahulikar_structure_2025}, we integrate over all frequencies the product of the analytic continuation of the photonic DOS and a window function which restricts to the frequency window $[\omega_0-\Delta\omega/2,\omega_0+\Delta\omega/2]$ for which we wish to generate a band gap. In describing the frequency window, we consider some central frequency $\omega_0$ and some bandwidth, or equivalently some band gap, $\Delta\omega$. We obtain the expression~\cite{bahulikar_structure_2025}
\begin{equation}
\tilde{\text{DOS}}_N(\omega_0,\Delta\omega) =  \text{Im}\left[\frac{\sum_{n = 0}^{N-1}\left(e^{i(\pi+2\pi n)/(2N)}\right)\text{DOS}'\left(\omega_0-\frac{\Delta\omega}{2}e^{i(\pi+2\pi n)/(2N)}\right)}{\csc\left(\frac{\pi}{2N}\right)}\right].\label{eq:finalresidue}
\end{equation}
In Eq. (\ref{eq:finalresidue}), $N$ is half the number of poles of the window function, which will serve as the number of frequencies at which we will sample the band gap. The quantity $\text{DOS}'\left(\omega_0-\frac{\Delta\omega}{2}e^{i(\pi+2\pi n)/(2N)}\right)$ is the analytic continuation of the photonic DOS evaluated at the prescribed complex frequency corresponding to a pole of the window function~\cite{bahulikar_structure_2025}. This analytic continuation of the photonic DOS is given by~\cite{bahulikar_structure_2025}
\begin{equation}
\text{DOS}'(\omega) = -\frac{6}{\pi}\sum_j\int\int\mathbf{J}_{j,\mathbf{r}'}^*(\mathbf{r})\cdot\mathbf{E}_{j,\mathbf{r}'}(\omega,\mathbf{r})\text{d}\mathbf{r}\text{d}\mathbf{r}'.
\end{equation}
For TE polarization, we sum over polarizations in the $x$ and $y$ directions, corresponding to unit vectors $\hat{e}_x$ and $\hat{e}_y$, given that any in-plane electric field can be decomposed into a field pointing along the $x$ direction and a field pointing along the $y$ direction. One also needs to ensure that a numerically stable approach is employed for taking gradients, which can be accomplished by replacing all the forward and backward finite differences with central differences in the ceviche code~\cite{hughes_forward-mode_2019}. In implementing the equations, the $+i\omega t$ time convention of the ceviche code must also be considered~\cite{hughes_forward-mode_2019}. We employ the NLopt package~\cite{johnson_nlopt_2007} with the Method of Moving Asymptotes (MMA) algorithm~\cite{svanberg_class_2002} for our optimization. Our innovation~\cite{bahulikar_structure_2025} is to reduce the computational demand of computing the analytic continuation of the photonic DOS by using a single uniform source covering all of the design region instead of a number of point dipole sources equal to the number of grid points in the design region. As described in our earlier work~\cite{bahulikar_structure_2025}, our uniform-source approach can be cast into full Brillouin zone or $\Gamma$-point formalisms, depending on whether or not the electric fields are expressed as Bloch states with non-zero wavevectors $\mathbf{k}$.

\section{Results} \label{sec:results}
\subsection{Photonic-crystal behavior as a function of system size and grid-point resolution}\label{sec:systemsize}
\begin{figure*}[ht!] 
\centering
\includegraphics[width=0.8\textwidth]{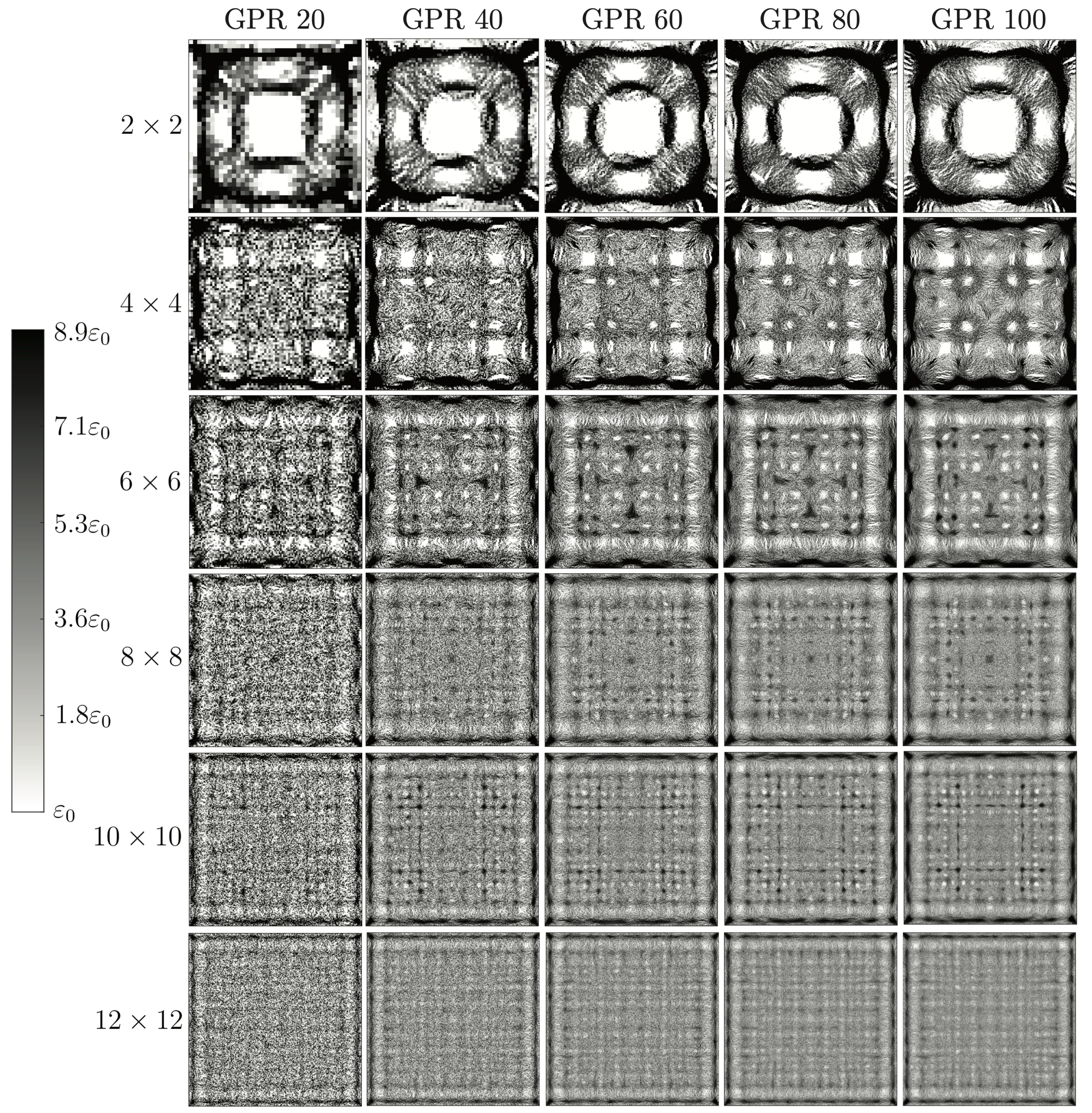}
\caption{Behavior of the results of our optimization algorithm as the grid-point resolution (GPR) and design-region size are increased. Structures are displayed at 5000 iterations. The value $N=10$ was employed. The color bar indicates the value of the permittivity $\epsilon(\mathbf{r})$ at the various point in the design region.} 
\label{fig:TE_sizes}
\end{figure*}

In the $\Gamma$-point formulation of our uniform-source approach, we investigate the photonic crystal behavior for transverse-electric (TE) polarization as a function of system size and grid-point resolution (GPR) (see Fig. \ref{fig:TE_sizes}). Here we use the scalar wave equation for the magnetic field implemented in the ceviche code~\cite{hughes_forward-mode_2019}, which has the form
\begin{align}
&\left(\mathbf{\nabla}\times\frac{1}{\epsilon(\mathbf{r})}\mathbf{\nabla}\times~ -~ \mu(\mathbf{r})\omega^2\right)\mathbf{H}_{x,y} = \nabla\times\left(\frac{\hat{e}_{x,y}}{\epsilon(\mathbf{r})V_{pc}}\right)\\ &\implies -\nabla\cdot\left(\frac{1}{\epsilon(\mathbf{r})}\nabla H_{x,y}\right)- \mu(\mathbf{r})\omega^2H_{x,y} = \left(\nabla\times\left(\frac{\hat{e}_{x,y}}{\epsilon(\mathbf{r})V_{pc}}\right)\right)_z.
\end{align}
Above, $V_{pc} = a^2$ is the volume of the primitive unit cell and $a$ is the lattice constant of the primitive unit cell, where a square primitive unit cell is assumed. The subscript $x,y$ implies the existence of two separate equations, one for the $x$ polarization and one for the $y$ polarization. We have also noted that $\nabla\left(\frac{1}{\epsilon(\mathbf{r})}\nabla \cdot \mathbf{H}_{x,y}\right) = 0$ since the sources $\nabla\times\left(\frac{\hat{e}_{x}}{\epsilon(\mathbf{r})V_{pc}}\right)$ and $\nabla\times\left(\frac{\hat{e}_{y}}{\epsilon(\mathbf{r})V_{pc}}\right)$ are polarized in the $z$ direction (derivatives are taken in the plane perpendicular to the $z$ direction). The polarization of the sources implies that $\mu(\mathbf{r})\omega^2\left(\mathbf{H}_{x,y}\right)_x = \mu(\mathbf{r})\omega^2\left(\mathbf{H}_{x,y}\right)_y = 0$ so that we can write the magnetic fields as the scalars $H_{x,y} = \left(\mathbf{H}_{x,y}\right)_z$. For comparison with the literature~\cite{joannopoulos_photonic_2011}, we employed a central frequency $\omega_0 = 0.4\cdot2\pi c/a$, a band gap $\Delta\omega = \omega_0/10$, allowed the permittivity $\epsilon(\mathbf{r})$ to vary between $\epsilon_0$ and $8.9\epsilon$, and set the permeability $\mu(\mathbf{r})$ equal to $\mu_0$. All results shown in this manuscript employ $\Delta\omega = \omega_0/10$. Each grid point in the design region was randomly initialized to a permittivity value between $\epsilon_0$ and $8.9\epsilon_0$. We set $a=1$, leveraging the scale invariance of Maxwell's equations~\cite{joannopoulos_photonic_2011}, and use $c$, $\mu_0$, and $\epsilon_0$ for the speed of light, the permeability, and the permittivity of free space, respectively. The GPR is therefore the number of grid points employed to resolve a length $a=1$.  We find that as the system size increases, the designs approach dielectric veins arranged in a square lattice~\cite{joannopoulos_photonic_2011}. For the $10\times10$ and $12\times12$ design region size and a GPR of 100 with $N=10$, we perform a Fourier transform of the optimized design at 5000 iterations. We find that the location of the four highest maxima of the absolute value of the transform that occur at positions other than the zero vector are at $\mathbf{k}_1 = 7\cdot 2\pi/(10a)\hat{e}_y$, $\mathbf{k}_2 = -7\cdot 2\pi/(10a)\hat{e}_y$, $\mathbf{k}_3 = -7\cdot 2\pi/(10a)\hat{e}_y$, $\mathbf{k}_2 = 7\cdot 2\pi/(10a)\hat{e}_y$ for the $10\times10$ design-region size. For the $12\times12$ design-region size, the maxima are located at $\mathbf{k}_1 = -12\cdot 2\pi/(12a)\hat{e}_y$, $\mathbf{k}_2 = 12\cdot 2\pi/(12a)\hat{e}_y$, $\mathbf{k}_3 = 12\cdot 2\pi/(12a)\hat{e}_y$, $\mathbf{k}_2 = -12\cdot 2\pi/(12a)\hat{e}_y$.  Thus, for the $10\times10$ and $12\times12$ design-region sizes, the structures have the underlying symmetry of the square lattice with the lattice constant $a_{10}' = \frac{10}{7}a$ and $a_{12}' = a$, respectively.

\begin{figure*}[ht!] 
\centering
\includegraphics[width=0.8\textwidth]{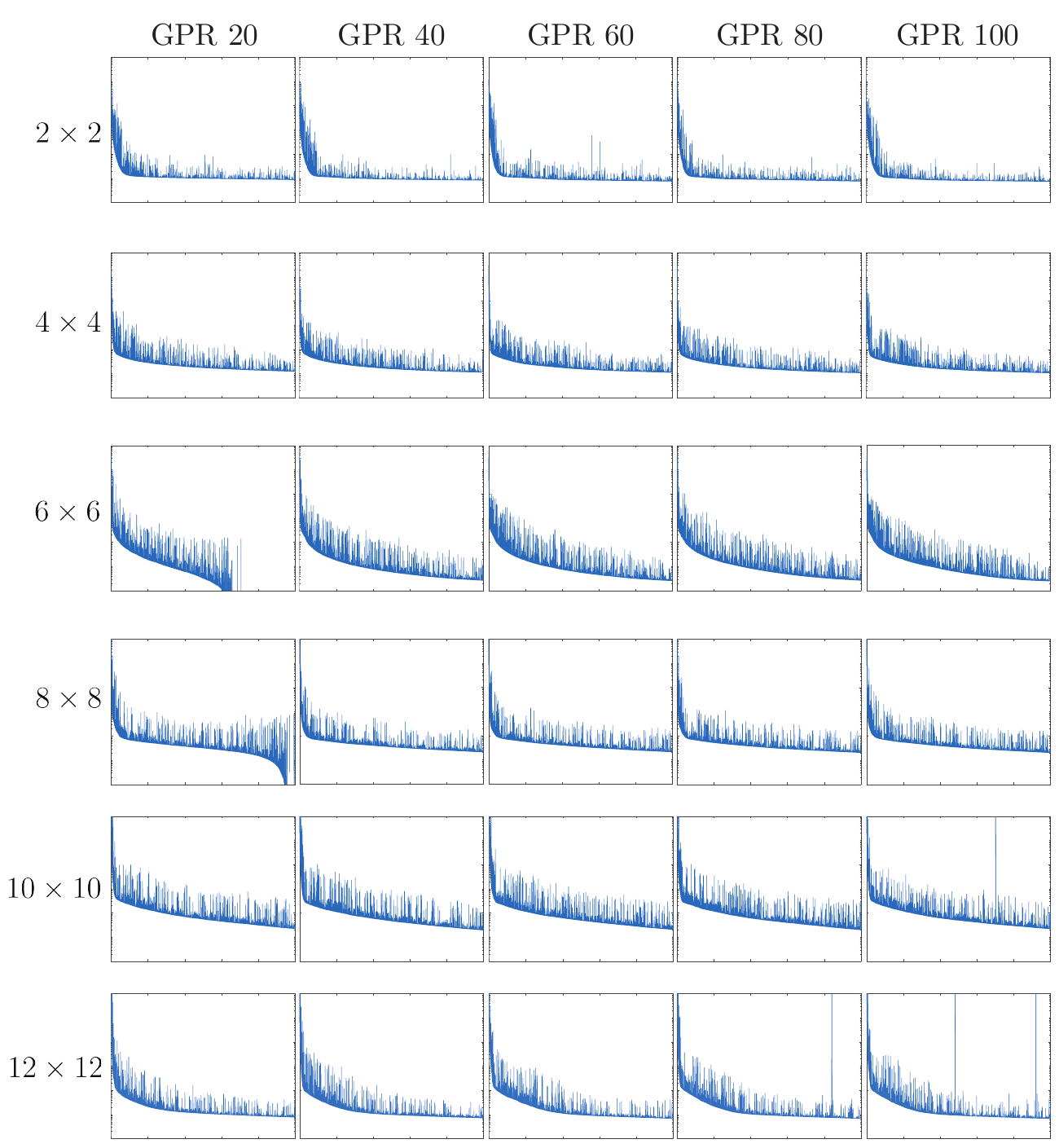}
\caption{Convergence of our objective $\tilde{\text{DOS}}_{10}(\omega_0,\Delta\omega)$ normalized by the corresponding quantity for vacuum ($\epsilon(\mathbf{r}) = \epsilon_0$)  in the entire design region as a function of the number of iterations for various sizes of the design region and various GPR values. The abscissae of the plots show the interval from 0 to 5000 iterations on a linear scale and the ordinate axes show the interval from $10^{-6}$ to 1 on a logarithmic scale.} 
\label{fig:TE_sizes_DOS}
\end{figure*}

As we discussed previously~\cite{bahulikar_structure_2025}, the Bragg length $L_B$, which captures the characteristic length for light attenuation within a photonic crystal and can be approximated by~\cite{hasan_finite_2018,neve-oz_bragg_2004}
 \begin{equation}
L_B \approx \frac{2d}{\pi}\frac{\omega_0}{\Delta\omega}. \label{eq:bragglength}
\end{equation}
informs the trend toward known photonic-crystal structures. The quantity $d$ in Eq. (\ref{eq:bragglength}) is the smallest distance for a set of crystal planes. In our case, for the $10\times10$ design-region size we find 
\begin{equation}
L_B \approx \frac{2d}{\pi}\frac{\omega_0}{\Delta\omega} = \frac{2a_{10}'}{\pi}\frac{\omega_0}{\omega_0/10} \approx 9. 
\end{equation}
By contrast, for the $12\times12$ design-region size we find 
\begin{equation}
L_B \approx \frac{2d}{\pi}\frac{\omega_0}{\Delta\omega} = \frac{2a_{12}'}{\pi}\frac{\omega_0}{\omega_0/10} \approx 6. 
\end{equation}
In order to consider structures that are comfortably within the photonic crystal regime, we employ the $12\times12$ design region size in investigating the behavior of the optimized designs as a function of $N$ in Section \ref{sec:Nbehavior} below. We show the convergence of our objective $\tilde{\text{DOS}}_{10}(\omega_0,\Delta\omega)$ normalized by the objective evaluated for vacuum ($\epsilon(\mathbf{r}) = \epsilon_0$) throughout the entire design region in Fig. \ref{fig:TE_sizes_DOS} for the parameter regimes in Fig. \ref{fig:TE_sizes}. For a GPR of 20, the $6\times6$ as well as the $8\times8$ design region sizes exhibit numerical instabilities, likely due to the closeness of the linear dimension of those design regions to $L_B$ evaluated for the $12\times12$ design region size.

\subsection{Photonic crystal behavior as a function of the number of sampled frequencies, $N$}
\begin{figure*}[ht!] 
\centering
\includegraphics[width=0.8\textwidth]{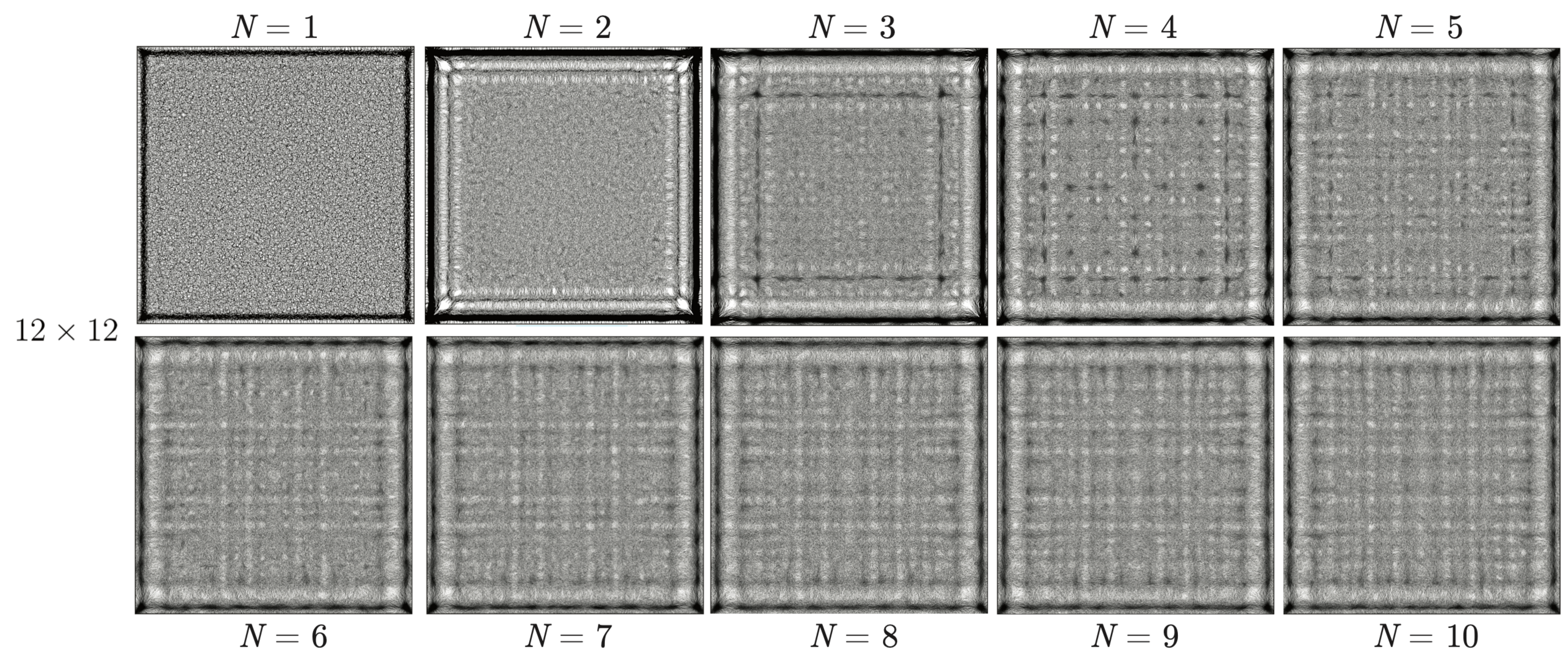}
\caption{Designs optimized for various $N$ with a GPR of 100 and employing a $12\times12$ design-region size. All designs completed 5000 iterations with the exception of the design for $N= 1$ where the optimization terminated after 444 iterations and is not well defined~\cite{bahulikar_structure_2025}. The permittivity follows the color scheme described by the color bar in Fig. \ref{fig:TE_sizes}.} 
\label{fig:TE_N}
\end{figure*}

\begin{figure*}[ht!] 
\centering
\includegraphics[width=0.8\textwidth]{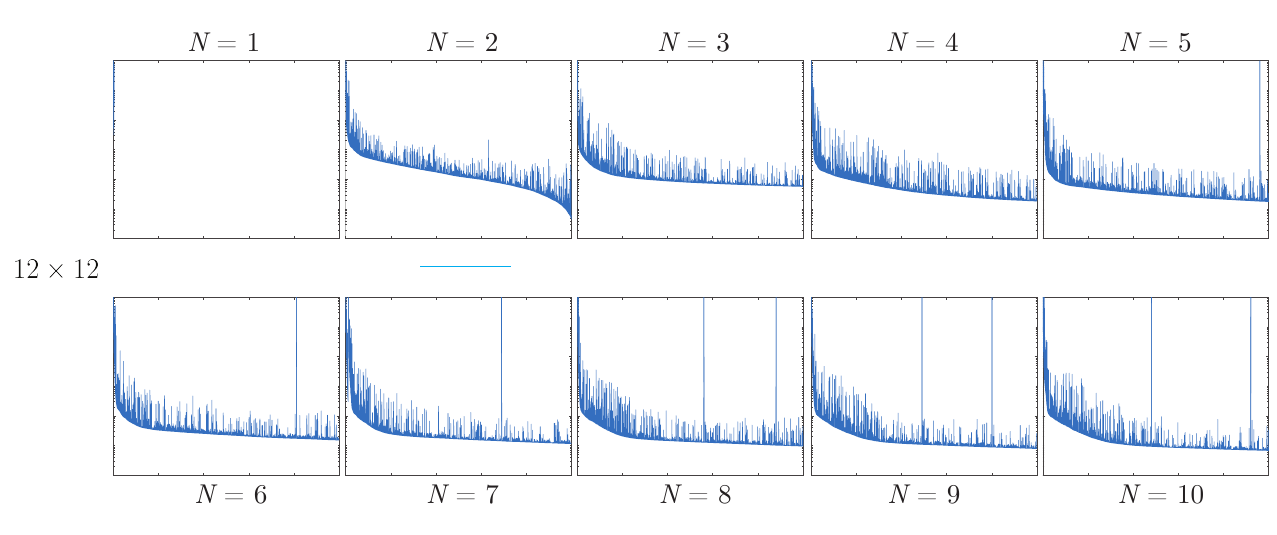}
\caption{Convergence of our objective $\tilde{\text{DOS}}_{N}(\omega_0,\Delta\omega)$ for various $N$ normalized by the corresponding quantity for vacuum in the entire design region as a function of the number of iterations for a $12\times12$ design-region size and a GPR of 100. The horizontal axes show the interval from 0 to 5000 iterations on a linear scale and the vertical axes show the interval from $10^{-6}$ to 1 on a logarithmic scale.} 
\label{fig:TE_N_DOS}
\end{figure*}

In this section, we consider the photonic crystal behavior as a function of $N$, again for TE polarization using the scalar wave equation for the magnetic field. We use random initialization of the design regions with permittivity values confined to the range $\frac{\epsilon(\mathbf{r})}{\epsilon_0} \in [1,8.9]$. As detailed in our earlier work~\cite{bahulikar_structure_2025}, the minimum $N$ to observe photonic crystal behavior is given by $N_\text{min} \approx \text{GPR}\cdot \frac{\Delta \omega}{ck_\text{min}} \approx 4$ for a GPR value of 100, $\Delta\omega = \left(0.4\cdot 2\pi c/a\right)/10$, and $k_{\text{min}} = 12\cdot 2\pi/(12a) = 2\pi/a$, which is consistent with the results observed in Fig. \ref{fig:TE_N}. We show convergence of our objective $\tilde{\text{DOS}}_{N}(\omega_0,\Delta\omega)$ for various $N$ in Fig. \ref{fig:TE_N_DOS}, where each evaluation of $\tilde{\text{DOS}}_{N}(\omega_0,\Delta\omega)$ is normalized by the corresponding quantity when the entire design region is filled with vacuum. The convergence of the objective function improves steadily with $N$, with the exception of $N= 1$ and $N=2$, which both became numerically unstable and exhibited negative $\tilde{\text{DOS}}_{N}(\omega_0,\Delta\omega)$ values.  

\label{sec:Nbehavior}

\subsection{Structure-adaptive nature of our approach}
\begin{figure*}[ht!] 
\centering
\includegraphics[width=0.99\textwidth]{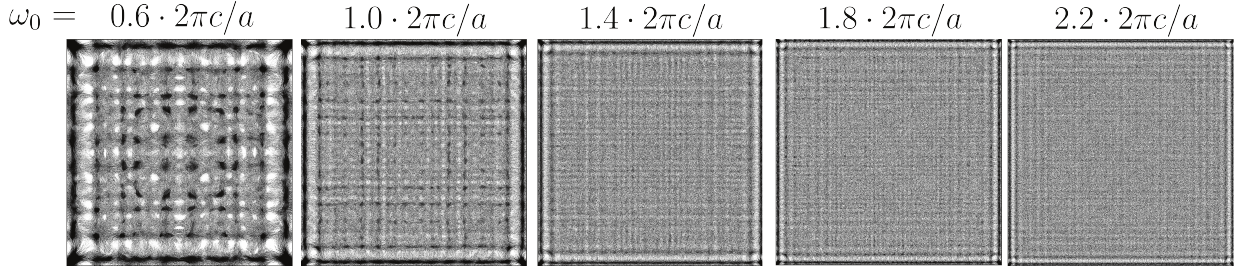}
\caption{Structures optimized for 2500 iterations for various $\omega_0$ in $10\times10$ design regions with a GPR of 100 and $N = 10$. TE polarization was used and $\epsilon(\mathbf{r})$ was confined to the range $\frac{\epsilon(\mathbf{r})}{\epsilon_0} \in [1,3.1]$. The darker colors correspond to larger permittivity values and the lighter colors to smaller permittivity values with the extremal permittivity values having the same coloring as the extremal permittivity values in Fig. \ref{fig:TE_sizes} and the remaining permittivity values adhering to the same scaling as in Fig. \ref{fig:TE_sizes}.}
\label{fig:varyomega}
\end{figure*}

\begin{figure*}[ht!] 
\centering
\includegraphics[width=0.99\textwidth]{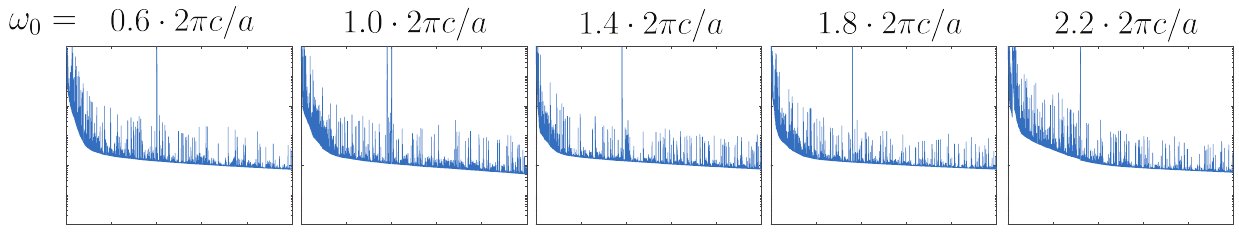}
\caption{Convergence of $\tilde{\text{DOS}}_{10}(\omega_0,\Delta\omega)$ for the designs displayed in Fig. \ref{fig:varyomega}. The quantity $\tilde{\text{DOS}}_{10}(\omega_0,\Delta\omega)$ is normalized by the vacuum value. The horizontal axes cover the range from 0 to 2500 on a linear scale, while the vertical axes cover the range from $10^{-6}$ to 1 on a logarithmic scale.}
\label{fig:varyomegaconv}
\end{figure*}

\begin{figure*}[ht!] 
\centering
\includegraphics[width=0.99\textwidth]{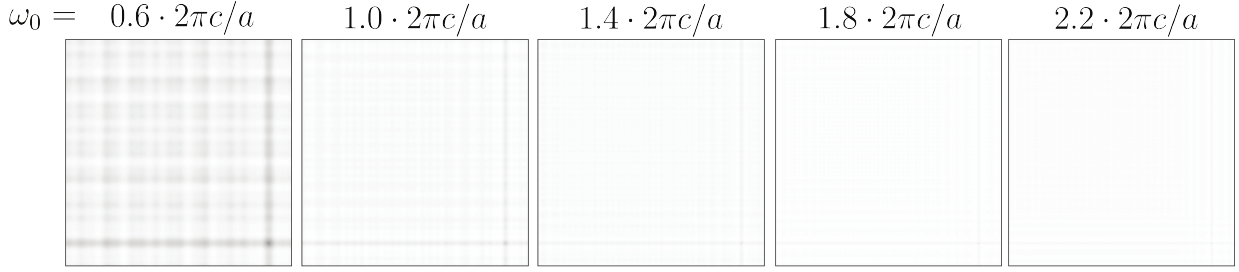}
\caption{Two-dimensional autocovariance functions for the designs displayed in Fig. \ref{fig:varyomega}.}
\label{fig:varyomegaautocovar}
\end{figure*}

We investigate the structure-adaptive nature of our approach by first considering different values of $\omega_0$ with $\frac{\epsilon(\mathbf{r})}{\epsilon_0} \in [1,3.1]$ for designs in square supercells optimized using the scalar wave equation for the magnetic field. The designs for $\frac{\epsilon(\mathbf{r})}{\epsilon_0} \in [1,3.1]$ are shown in Fig. \ref{fig:varyomega} using $N=10$, a $10\times10$ design region, a GPR of 100, and random initialization of the permittivity values within the prescribed range. The corresponding convergence of the objective function $\tilde{\text{DOS}}_{N}(\omega_0,\Delta\omega)$ for the various $\omega_0$ is shown in Fig. \ref{fig:varyomegaconv}. We subdivide the design regions into 10 equal parts along each dimension (100 equal parts in total) and apply a Fourier transform to each of the parts. We find that the maxima of the Fourier transform that occur at non-trivial positions partition the design into regions where primitive lattice vectors are generally oriented along the $x$ and $y$ directions. We find on the order of one part per supercell where 6-fold symmetry is observed to within an angle of 5$^\circ$ for $\omega_0 \geq 1.4\cdot 2\pi c/a$ and none for $\omega_0 < 1.4\cdot 2\pi c/a$. We compute also the autocovariance function in 2D (see Fig. \ref{fig:varyomegaautocovar}), where the decaying signal amplitude reflects the disorder in the system~\cite{marchetti_decoding_2004}. 
 
\begin{figure*}[ht!] 
\centering
\includegraphics[width=0.99\textwidth]{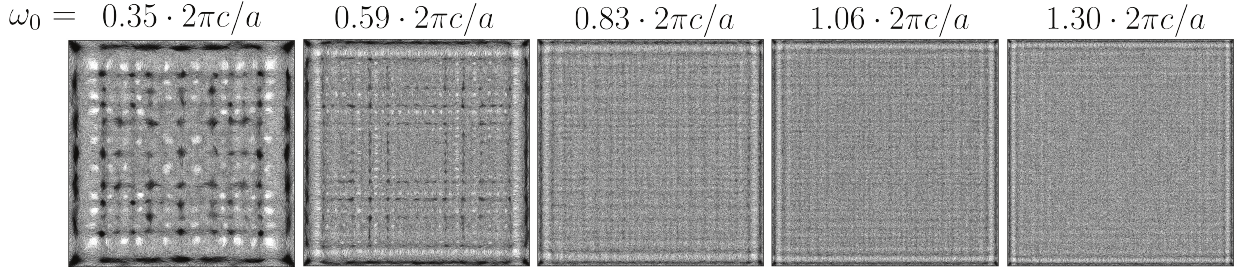}
\caption{Designs optimized for 2500 iterations for various $\omega_0$ consistent with the values in Fig. \ref{fig:varyomega} using the scale invariance of Maxwell's equations in $10\times10$ design regions with a GPR of 100 and $N=10$. We employed TE-polarized sources in generating the designs and $\epsilon(\mathbf{r})$ was confined to the range $\frac{\epsilon(\mathbf{r})}{\epsilon_0} \in [1,8.9]$. The color scheme for these designs is the same as in Fig. \ref{fig:TE_sizes}.}
\label{fig:varyomega8d9}
\end{figure*}

\begin{figure*}[ht!] 
\centering
\includegraphics[width=0.99\textwidth]{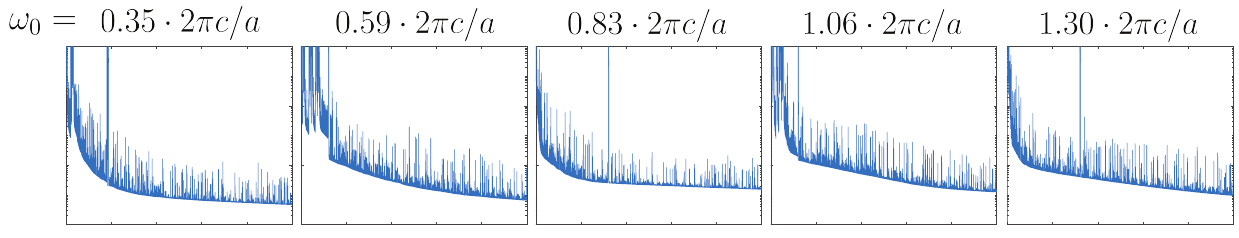}
\caption{Above is displayed the convergence of the quantity $\tilde{\text{DOS}}_{10}(\omega_0,\Delta\omega)$ normalized by the vacuum value for the designs shown in Fig. \ref{fig:varyomega8d9}. The range from 0 to 2500 is covered by the horizontal axes on a linear scale, while the range from $10^{-6}$ to 1 is covered on the vertical axes on a logarithmic scale.}
\label{fig:varyomega8d9conv}
\end{figure*}

\begin{figure*}[ht!] 
\centering
\includegraphics[width=0.99\textwidth]{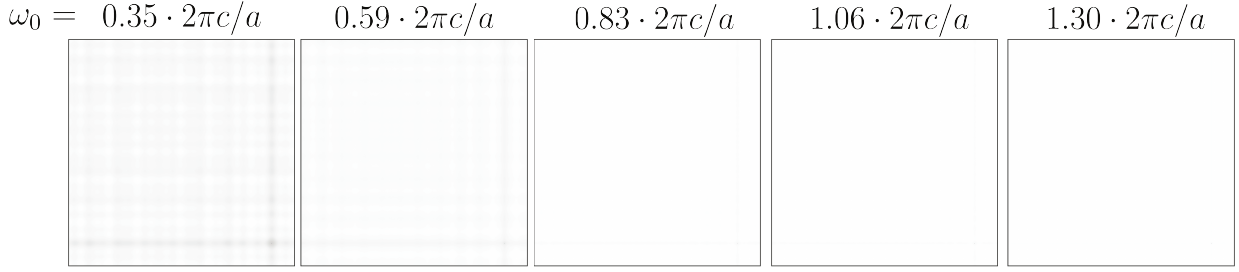}
\caption{The autocovariance functions in 2D for the designs shown in Fig. \ref{fig:varyomega8d9} are displayed above.}
\label{fig:varyomega8d9autocovar}
\end{figure*}

\begin{figure*}[ht!] 
\centering
\includegraphics[width=0.5\textwidth]{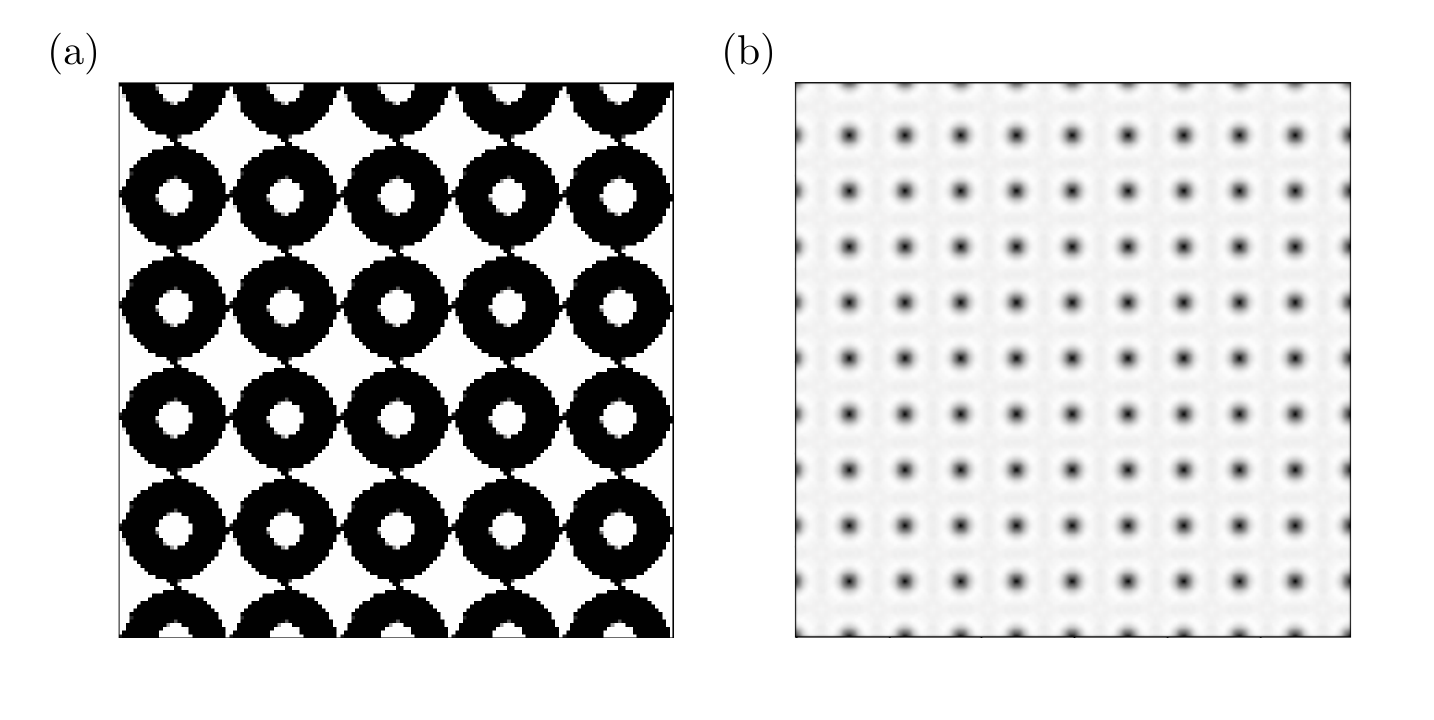}
\caption{Sample crystal with perfect periodicity (a) and corresponding autocovariance function (b). }
\label{fig:crystal_autocovar}
\end{figure*}

We next consider $\frac{\epsilon(\mathbf{r})}{\epsilon_0} \in [1,8.9]$ for structures in square $10\times10$ supercells optimized using the scalar wave equation for the magnetic field with $N=10$ and a GPR of 100 for various $\omega_0$ in investigating the structure-adaptive nature of our approach. We employ the scale invariance of Maxwell's equations~\cite{joannopoulos_photonic_2011} in assigning the $\omega_0$ values such as to achieve correspondence with the values in Fig. \ref{fig:varyomega}. The resulting designs are shown in Fig. \ref{fig:varyomega8d9}. The structures in the optimizations were again randomly initialized with permittivity values within the prescribed range. The convergence of $\tilde{\text{DOS}}_{N}(\omega_0,\Delta\omega)$ for the various $\omega_0$ is displayed in Fig. \ref{fig:varyomega8d9conv}. We see that for this larger permittivity value greater suppression of our objective function is obtained, which is consistent with known results~\cite{joannopoulos_photonic_2011,rechtsman_method_2009}. The Fourier analysis described above yields a partitioning of the design regions into cells with primitive lattice vectors that are generally oriented along the $x$ and $y$ directions. We find on the order of one cell per supercell where 6-fold symmetry is observed to within an angle of 5$^\circ$ for $\omega_0 < 1.30 \cdot 2\pi c/a$, which increases by nearly an order of magnitude for $\omega_0 =1.30 \cdot 2\pi c/a$.  We include as well the autocovariance function in 2D (see Fig. \ref{fig:varyomega8d9autocovar}), which displays decaying signal amplitude consistent with a disordered system. By contrast, we show in Fig. \ref{fig:crystal_autocovar} that the autocovariance function of a periodically repeating crystal does not exhibit a decaying signal.

\begin{figure*}[ht!] 
\centering
\includegraphics[width=0.9\textwidth]{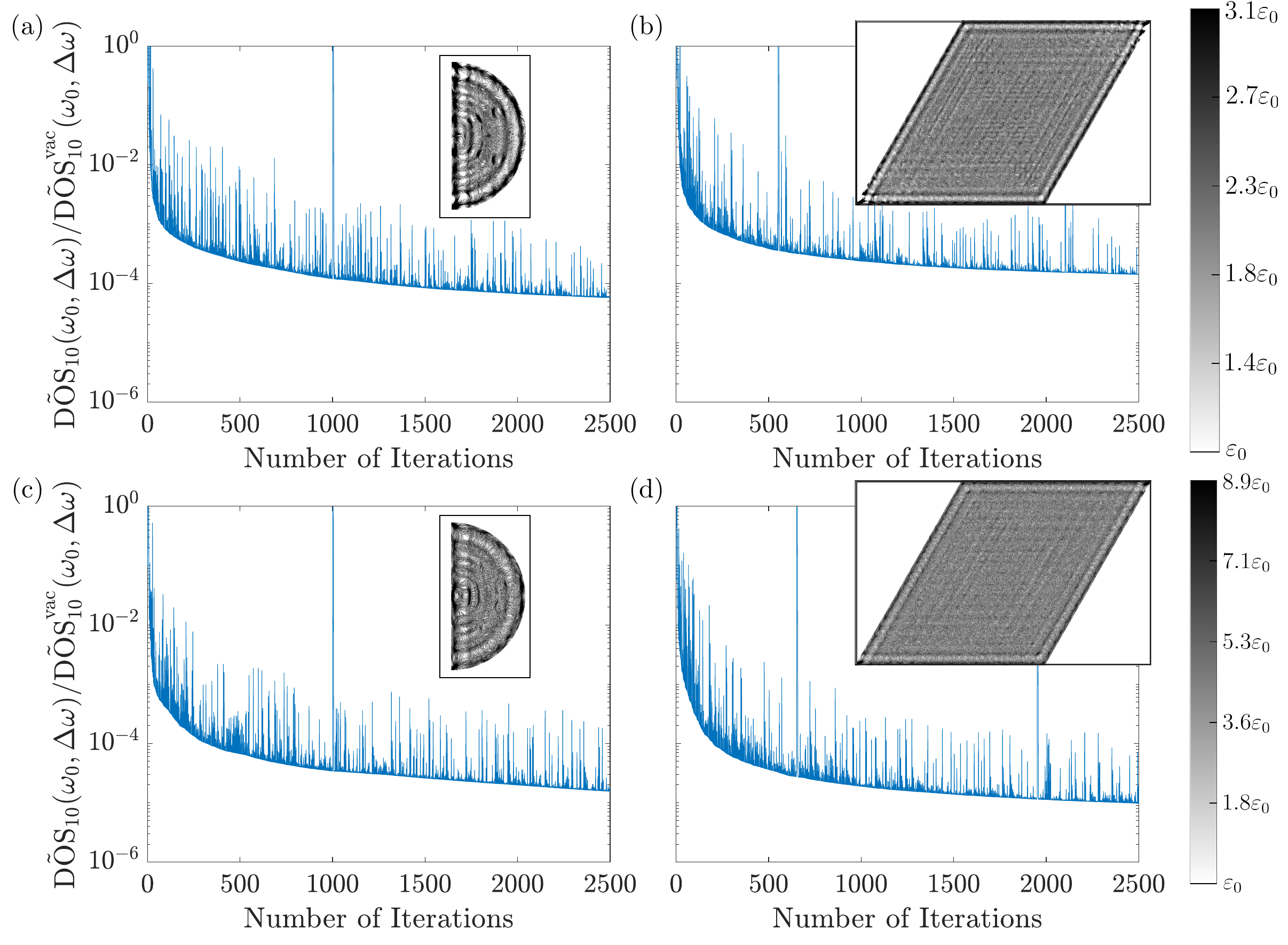}
\caption{Convergence of $\tilde{\text{DOS}}_{10}(\omega_0,\Delta\omega)$ normalized by the corresponding vacuum quantity is shown above with $\frac{\epsilon(\mathbf{r})}{\epsilon_0} \in [1,3.1]$ for a half circle (a) and for a supercell of a rhombus structure (b), and with $\frac{\epsilon(\mathbf{r})}{\epsilon_0} \in [1,8.9]$ for a half circle (c) and for a supercell of a rhombus structure in (d). }
\label{fig:various_designs_eps3d1_eps8d9}
\end{figure*}

We conclude this section with an explicit demonstration of the ability of our optimization algorithm to adapt to various design regions. We show in Fig. \ref{fig:various_designs_eps3d1_eps8d9} designs for a half circle and a supercell of a rhombus structure for permittivity values within the ranges $\frac{\epsilon(\mathbf{r})}{\epsilon_0} \in [1,3.1]$ for which $\omega_0 = 1.4 \cdot 2\pi c/a$ was employed and $\frac{\epsilon(\mathbf{r})}{\epsilon_0} \in [1,8.9]$ for which $\omega_0 = 0.83 \cdot 2\pi c/a$ was used, and with $N= 10$ and a GPR of 100. The designs were optimized using the scalar wave equation for the magnetic field. Performing the Fourier analysis described above for the supercells of a rhombus structure that were mapped into rectangular designs using periodic boundary conditions, we find that for $\frac{\epsilon(\mathbf{r})}{\epsilon_0} \in [1,3.1]$ the design region partitions into cells that obey 6-fold symmetry with no cells obeying 4-fold symmetry. For $\frac{\epsilon(\mathbf{r})}{\epsilon_0} \in [1,8.9]$, we find that the design region generally partitions into cells obeying 6-fold symmetry with on the order of one cell obeying 4-fold symmetry. This finding is consistent with the fact that 4-fold photonic-crystal structures have been found not to exhibit 10\% photonic band gaps for $\frac{\epsilon(\mathbf{r})}{\epsilon_0} \in [1,3.1]$~\cite{rechtsman_method_2009}. Again, larger permittivity values lead to greater suppression of our objective function, consistent with the literature\cite{joannopoulos_photonic_2011,rechtsman_method_2009}.

\subsection{Validity of our approach}
\begin{figure*}[ht!] 
\centering
\includegraphics[width=0.99\textwidth]{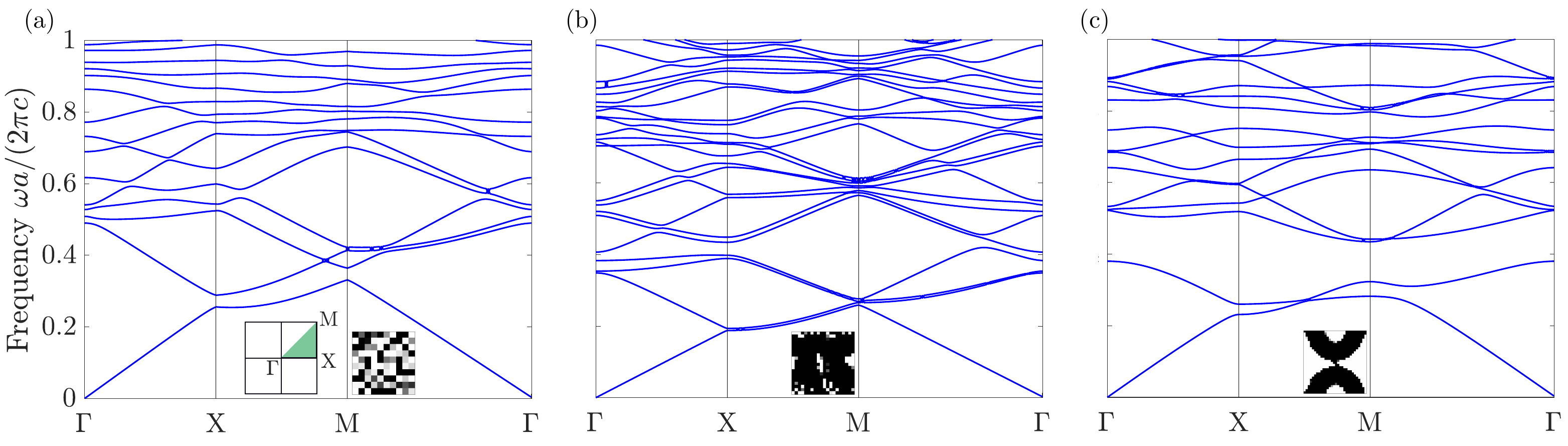}
\caption{TE bandstructures obtained for $1\times1$ unit cells with GPR values of 10 (a), 20 (b), and 30 (c). The bandstructure path is illustrated in (a). We employed $\omega_0 = 0.4\cdot 2\pi c/a$ and $\Delta \omega = \omega_0/10$ in generating the designs, which are inset in the bandstructure plots. The permittivity values were confined to the range $\frac{\epsilon(\mathbf{r})}{\epsilon_0} \in [1,8.9]$ so that the color scheme follows Fig. \ref{fig:TE_sizes}. }
\label{fig:bandstructures_unitcell}
\end{figure*}

\begin{figure*}[ht!] 
\centering
\includegraphics[width=0.99\textwidth]{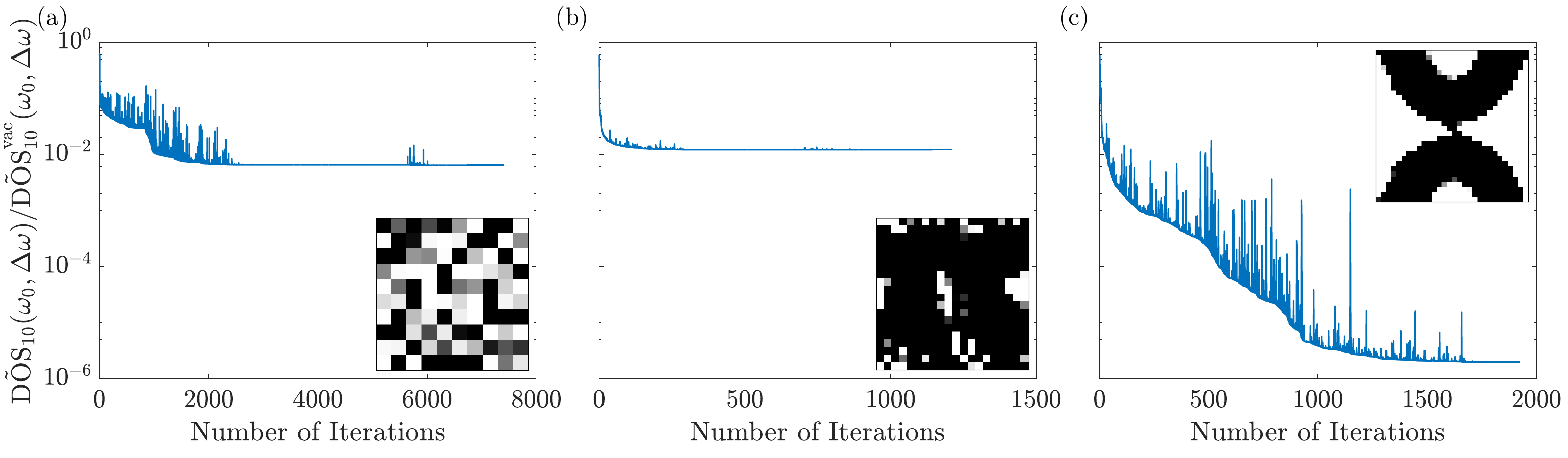}
\caption{Convergence of $\tilde{\text{DOS}}_{10}(\omega_0,\Delta\omega)$ normalized by the corresponding vacuum value for the designs in Fig. \ref{fig:bandstructures_unitcell}. We have again inset the designs in the plots. The abscissae of the plots cover the range from 0 to the total number of iterations for complete convergence on a linear scale an the ordinate axes are shown over the interval from $10^{-6}$ to 1.}
\label{fig:DOSconvergence_unitcell}
\end{figure*}

In order to validate our approach, we compute TE bandstructures for designs optimized with GPR values of 10, 20, and 30 using an initialization of alternating high permittivity and low permittivity stripes. In the optimizations, we use the vector electric field formulation presented in Eqs. (\ref{eq:MaxwellE}) and (\ref{eq:MaxwellopE}) for the electromagnetic wave equation. In order to compute the bandstructures, we solve the equation
\begin{equation}
\mathbf{\nabla}_{\mathbf{k}}\times\frac{1}{\epsilon(\mathbf{r})}\mathbf{\nabla}_{\mathbf{k}}\times\mathbf{u}^{\mathbf{H}}_{j,\mathbf{k}} = \mu(\mathbf{r})\omega_\mathbf{k}^2\mathbf{u}^{\mathbf{H}}_{j,\mathbf{k}},
\end{equation}
where 
\begin{equation}
\mathbf{\nabla}_\mathbf{k} = \mathbf{\nabla}+i\mathbf{k}
\end{equation}
for the wavevectors $\mathbf{k}$ along the bandstructure path in Fig. \ref{fig:bandstructures_unitcell} (a) and the magnetic-field eigenmodes $\mathbf{u}^\mathbf{H}_{j,\mathbf{k}}$.
We employ $N=10$ in generating the designs and use a $1\times1$ unit cell with periodic boundary conditions. In Fig. \ref{fig:bandstructures_unitcell}, we present the resulting designs from the stripes initializations. For a GPR of 10 and 20, no gap opens, while for a GPR of 30 we achieve a central frequency of $\omega_0 = 0.41\cdot 2\pi c/a$, in good agreement with our desired midgap frequency of $\omega_0 = 0.4\cdot 2\pi c/a$, and a gap of $\Delta\omega = 0.11 \cdot \omega_0$, which satisfies the optimization criterion that the gap be at least $\omega_0/10$. We observe significant suppression of $\tilde{\text{DOS}}_{10}(\omega_0,\Delta\omega)$ for the GPR value of 30 as shown in Fig. \ref{fig:DOSconvergence_unitcell}. 

\subsection{Applications}

\begin{figure*}[ht!] 
\centering
\includegraphics[width=0.99\textwidth]{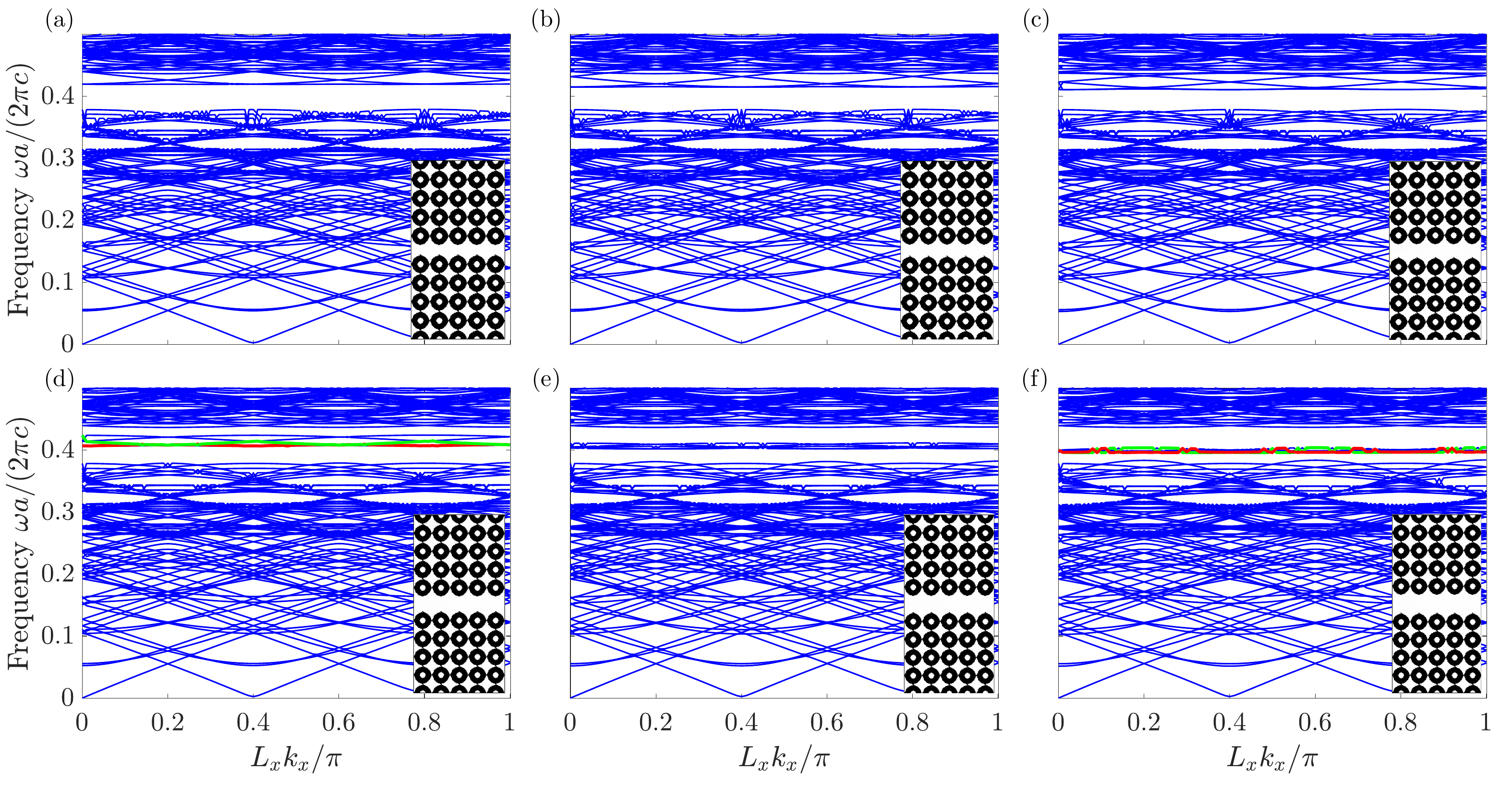}
\caption{Above, the TE bandstructures obtained for waveguide systems consisting of slabs separated by 0.5, 0.6, 0.7, 0.8, 0.9, and 1.0 in units of $a$, with horizontal dimension $L_x = 5a$ and periodic boundary conditions. All waveguides employed the unit-cell design with a GPR of 30 from Figs. \ref{fig:bandstructures_unitcell} (c) and \ref{fig:DOSconvergence_unitcell} (c). No additional optimization was performed for the unit cell designs placed in waveguide structures. The bands for which the electric field intensities have been calculated are illustrated in red and green in (d) and (f).} 
\label{fig:bandstructures_waveguide}
\end{figure*}

\begin{figure*}[ht!] 
\centering
\includegraphics[width=0.87\textwidth]{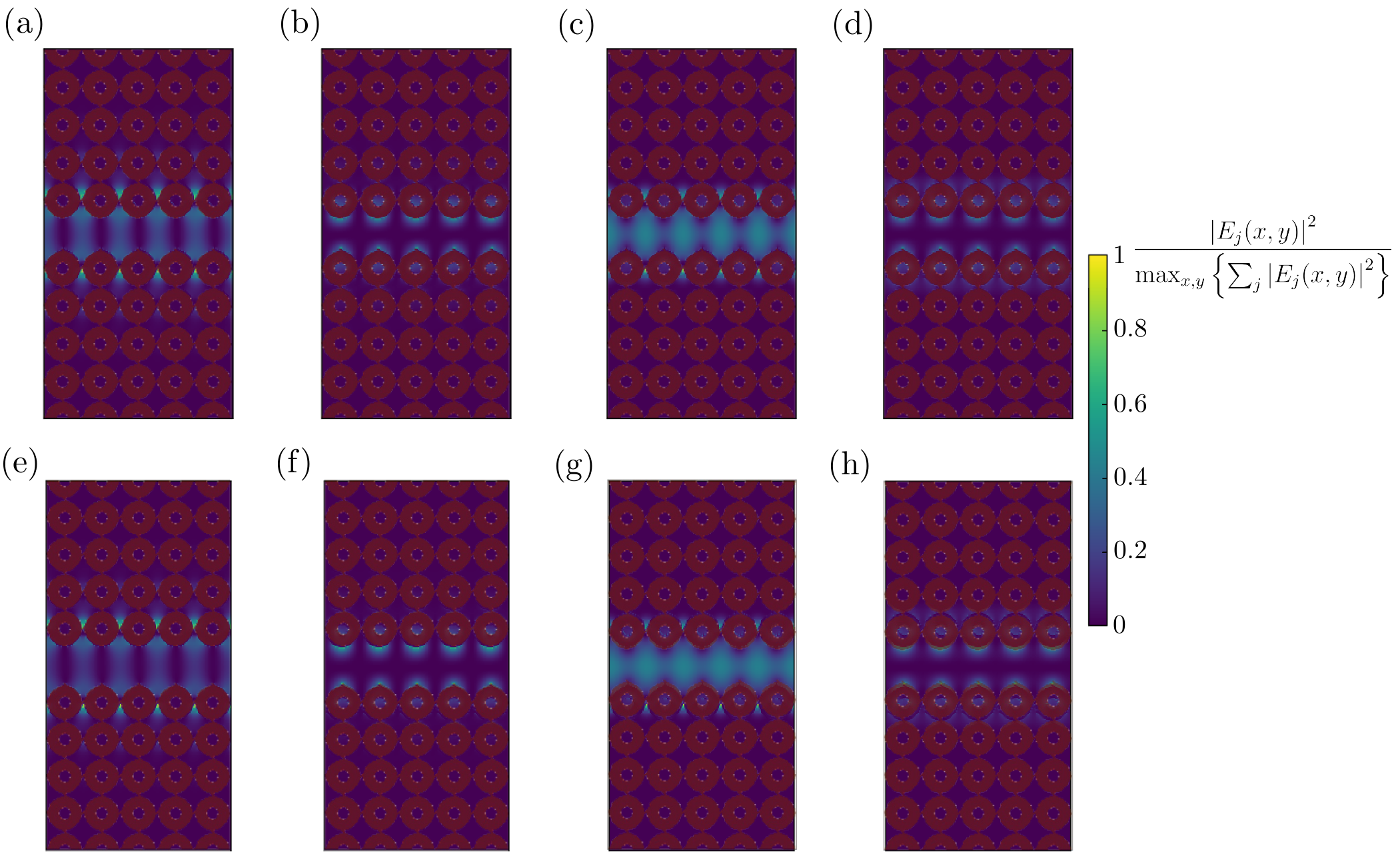}
\caption{The squared absolute values of the $x$ and $y$ components of the electric field normalized by the corresponding maximum electric field intensities are shown above. The $x$ component of the electric field for rightward traveling modes is shown in (a) and (e) for separations of 0.8 and 1.0, respectively, while the $y$ component of the electric field for rightward traveling modes is shown in (b) and (f) for separations of 0.8 and 1.0, respectively. For the leftward traveling modes the $x$ component of the electric field is shown in (c) for a separation of 0.8 and (g) for a separation of 1.0, while the $y$ component of the electric field is shown in (d) for a separation of 0.8 and (h) for a separation of 1.0. All modes were evaluated at $L_xk_x/\pi = 0.5$.}
\label{fig:Efield_waveguide}
\end{figure*}

We consider an application of our optimization results in the design of waveguides. For a GPR of 30, we show in Fig \ref{fig:bandstructures_waveguide} the bandstructure plots for systems consisting of two slabs with separations of 0.5, 0.6, 0.7, 0.8, 0.9, and 1.0 in units of $a$ constructed from the unit cell design in Figs. \ref{fig:bandstructures_unitcell} (c) and \ref{fig:DOSconvergence_unitcell} (c), and with horizontal dimension $L_x = 5a$. We determine the magnetic field eigenmodes at $L_xk_x/\pi = 0.5$ and employ Eq. (\ref{eq:HtoE}) to compute the corresponding $x$ and $y$ components of the electric field. These are displayed in Fig. \ref{fig:Efield_waveguide} for rightward traveling modes corresponding to a positive $\frac{\partial \omega}{\partial k_x}$ (the red bands in Fig. \ref{fig:bandstructures_waveguide} (d) and (f)) as well as for leftward traveling modes corresponding to a negative group velocity $\frac{\partial \omega}{\partial k_x}$ (the green bands in Fig. \ref{fig:bandstructures_waveguide} (d) and (f)). We compute the modes for a separation of 0.8 to aid in the identification of the rightward or leftward traveling modes for a separation of 1.0 where beats are observed in the bandstructure due to closely spaced frequencies. 

\subsection{Extension to inversion of bandstructures}
We note also that our framework readily generalizes to the solution of the problem of inverting a photonic bandstructure for a given crystal symmetry. Our framework could apply as well to solving the problem of inverting a phononic bandstructure for a particular crystal symmetry given the similarities in the structure of the equations for the two problems. In order to invert the photonic bandstructure, one would express the electric fields as Bloch states indexed by wavevectors $\mathbf{k}$. For each pair of $\mathbf{k}$ and $\omega$, the LDOS computed for a uniform source would be enhanced (suppressed) wherever a band should (should not) exist.
\label{sec:extension}

\section{Conclusion} \label{sec:conc}
In conclusion, we have presented a structure-adaptive approach for optimization for photonic band gaps with TE-polarized sources, which can treat frequency-dependent optical response. We cast our approach in formalisms based on $\Gamma$-point integration and integration over a full Brillouin zone and show that we can generate TE photonic crystals with desired band gaps. We also show that our approach can be extended to invert arbitrary bandstructures both for photonic systems and for phononic systems. Work treating complete photonic band gaps and extensions to three dimensions is forthcoming.

\section{Acknowledgements}
R.K.D. acknowledges financial support that made this work possible from Syracuse University College of Engineering and Computer Science funds. The authors also acknowledge that the work reported on in this paper was substantially performed using Zest, the Syracuse University research computing high-performance computing cluster.

\bibliography{refs_PC}
\end{document}